\documentclass[preprint]{aastex}
\usepackage{color}
\def\etal   {{\rm ~et al.,\,}}
\def\kms    {\ifmmode{{\rm ~km~s}^{-1}}\else{~km~s$^{-1}$}\fi}
\def\lsun   {\ifmmode{{\rm ~L}_\odot}\else{~L$_\odot$}\fi}
\def\msun   {\ifmmode{{\rm ~M}_\odot}\else{~M$_\odot$}\fi}
\def\red#1  {\textcolor{red}{#1}\ }
\shortauthors{Greenhill\etal}
\shorttitle{The Inner Parsec of the Circinus AGN}

\begin{document}
\font\cmss=cmss10 scaled 1200
\font\cmssbx=cmssbx10 scaled 1200

%\slugcomment{\tt DRAFT, 2002 Nov. 21, for submission to ApJ (Part 1)}

\title{A Warped Accretion Disk and Wide Angle Outflow in the Inner Parsec of the
Circinus Galaxy}

\author{
L. J. Greenhill,\altaffilmark{1}
R. S. Booth,\altaffilmark{2} \\
S. P. Ellingsen,\altaffilmark{3}
J. R. Herrnstein,\altaffilmark{1,4}
D. L. Jauncey,\altaffilmark{5} 
P. M. McCulloch,\altaffilmark{3} \\
J. M. Moran,\altaffilmark{1}
R. P. Norris,\altaffilmark{5}
J. E. Reynolds,\altaffilmark{5}
A. K. Tzioumis,\altaffilmark{5}
}

\altaffiltext{1}{Harvard-Smithsonian Center for Astrophysics, 60 Garden St, 
Cambridge, MA 02138 USA greenhill@cfa.harvard.edu, moran@cfa.harvard.edu}

\altaffiltext{2}{Onsala Space Observatory, 
Chalmers Institute of Technology, Onsala, S-43992 Sweden roy@oso.chalmers.se}

\altaffiltext{3}{University of Tasmania, School of Physics, 
GPO 252-21, Hobart, TAS 7001 Australia simon.ellingsen@utas.edu.au,
peter.mcculloch@utas.edu.au}

\altaffiltext{4}{Current address: Renaissance Technologies,
25 E. Loop Dr., Stony Brook, NY  11790 jrh@rentec.com} 

\altaffiltext{5}{Australia Telescope National Facility, P.O.
Box 76,  Epping, NSW 2121 Australia djauncey@atnf.csiro.au, rnorris@atnf.csiro.au,
jreynold@atnf.csiro.au, atzioumi@atnf.csiro.au }

\begin{abstract}

We present the first VLBI maps of H$_2$O maser emission ($\lambda 1.3$~cm) in the
nucleus of the Circinus Galaxy, constructed from data obtained with the Australia
Telescope Long Baseline Array.  The maser emission traces  a warped, edge-on
accretion disk between radii of $0.11\pm0.02$ and $\sim 0.40$ pc, as well as a
wide-angle outflow that extends up to $\sim 1$~pc from the estimated disk center. 
The disk rotation is close to Keplerian ($v\propto r^{-0.5}$), the maximum detected
rotation speed is $260\kms$, and the inferred central mass is
$1.7\pm0.3\times10^6\msun$.  The outflowing masers are irregularly distributed above
and below the disk, with relative outflow velocities up to $\sim \pm 160$\kms,
projected along the line of sight.  The flow probably originates closer than 0.1~pc
to the central engine, possibly in an inward extension of the accretion disk, though
there is only weak evidence of rotation in the outward moving material.  We observe
that the warp of the disk appears to collimate the outflow and to fix the extent of
the ionization cone observed on larger angular scales.  This study provides the
first direct evidence (i.e., through imaging) of dusty, high-density, molecular
material in a nuclear outflow $<1$~pc from the central engine of a Seyfert galaxy,
as well as the first graphic evidence that warped accretion disks can channel
outflows and illumination patterns in AGN. We speculate that the same arrangement,
which in some ways obviates the need for a geometrically thick, dusty torus, may
apply to
 other type-2 AGN.

\end{abstract}

\keywords{galaxies: active --- galaxies: individual (Circinus galaxy) ---
galaxies: Seyfert --- ISM: jets and outflows --- ISM: molecules --- masers}

\section{Introduction}

At a distance of $4.2\pm0.8$ Mpc \citep{freeman77}, the Circinus galaxy hosts
one of the nearest Seyfert~II nuclei.  The bolometric luminosity is $\sim
4\times10^{43}$ ergs\,s$^{-1}$, $\sim 10\%$ of which is contributed by star
formation in the inner $\sim 100$~pc \citep{moorwood96}.  The central engine
radiates $2\times10^{42}$ erg s$^{-1}$ (2-10 keV) and is obscured by a n$_{\rm
H}\sim4\times10^{24}$ cm$^{-2}$ column at energies lower than $\sim 10$ keV
\citep{matt99}.  Reflection by cool material dominates the X-ray spectrum,
which includes a prominent ($\sim 2$~keV equivalent width) Fe K$\alpha$
fluorescence line \citep{matt96}.  The optical spectrum includes coronal lines
from photoionized gas \citep{oliva94} and broad polarized H$\alpha$ emission,
which is $\sim 3300$\kms~wide and betrays the existence of an obscured broad
line region \citep{oliva98}.  The mass of the central engine has been
uncertain.  At $2.2\mu$m wavelength, the central non-stellar source is
$<1.5$~pc in radius, and its dynamical mass is
$<4\times10^6$\msun\citep{maiolino98}.  Although massive, the central engine
constitutes a small fraction of the nuclear mass, which including stars and gas 
amounts to $3.2\pm0.8\times10^8\msun$ inside a radius of 140~pc
\citep{curran98}.

The active galactic nucleus (AGN) of the Circinus galaxy drives a visible
nuclear outflow on kpc scales, the position angle of which suggests an
orientation for the central engine. \citet{elmouttie98} observe bipolar radio
lobes, which lie at position angles (PA) of $\sim115$ and
$295^\circ\pm5^\circ$. These are aligned with the minor axes of the galactic
H\,I disk (PA$=300^\circ\pm 5^\circ$) \citep{freeman77} and nuclear $^{12}$CO
ring (PA$=304^\circ\pm 4^\circ$), which lies between radii of 140 and 600~pc
and is inclined by $78^\circ\pm 1^\circ$ \citep{curran98}.  Circinus also
exhibits a kpc-scale, one-sided optical ionization cone \citep{marconi94} with
an opening angle of $\sim100^\circ$ and a mean PA$\sim 295^\circ$,
approximately along the outflow axis.  The ionization cone comprises knots and
radial filaments whose kinematics are indicative of outward motion, possibly
in the form of bullet-like ejecta, at speeds on the order of 100 to 200
\kms~\citep{vb97, elmouttie98, wilson00}.

There is good evidence that extragalactic H$_2$O masers ($\lambda 1.3$~cm) in
NGC\,4258, NGC\,1068, and other galaxies \citep[e.g.,][]{miyoshi95, gg97} trace
edge-on, warped accretion disks bound by central engines $\ga10^6$ M$_\odot$.
Spectral-line interferometer data with milliarcsecond resolution
straightforwardly provide accurate measurements of rotation curves, central engine
masses, warp shapes, and disk orientations. The maser spectra in the systems that are
the best examples exhibit complexes of ``high-velocity'' emission symmetrically
bracketing ``low-velocity'' emission that is close to the systemic velocity of
the host galaxy.   In Circinus, H$_2$O maser emission has been known for over
20 years \citep{gw82}, but only the relatively recent discoveries of spectral
features blueward of the systemic velocity \citep{nakai95, braatz96,
greenhill97} have led to speculation that the masers in Circinus also trace an
accretion disk.

We present the first maps of the Circinus H$_2$O maser source and report the
detection of an accretion disk, bearing out the earlier speculation. We have
also found that a fraction of the maser emission originates outside the disk,
in what appears to be a wide-angle outflow within $\sim 1$~pc of the central
engine and aligned with the ionization cone.  The discovery of this
qualitatively new phenomenon introduces a fresh element to the study of AGN,
because the locations and velocities of dense non-disk material in 
such close proximity to a central
engine can be mapped for the first time.  As a result, it is possible to test
in new ways proposed mechanisms for the driving and collimation of outflows in
AGN \citep[e.g.,][and references therein]{bottorff97}.

In Sections~2, 3, and 4, we summarize our Very Long Baseline Interferometer
(VLBI) observations, post-correlation analysis, and proposed model of a warped
disk and outflow.  In Section~5 we quantify the model and evaluate
ramifications for collimation and driving mechanisms, and discuss the relation of the
warped disk to the outflow and the ionization cone.

\section{Observations}

We observed the $6_{16}-5_{23}$ transition of H$_2$O ($\nu_{\rm rest}= 22235.08$
MHz) toward Circinus three times in 1997 and 1998 (Table\,1) with the
Australia Telescope Long Baseline Array (LBA)\footnote{The Australia Telescope Long
Baseline Array is part of the Australia Telescope, which is funded by the
Commonwealth of Australia for operation as a National Facility managed by CSIRO.},
which provided baselines of $\sim 200$ to 1000 km.  Table\,2 summarizes
the characteristics of the individual antennas.  Typical tracks were 18 h at Mopra
and Hobart, 12 h at Tidbinbilla, and 14 h at Parkes. Approximately every 1.5~h we
observed a continuum source for $\sim 0.5$~h (PKS\,0537-448, PKS\,1424-418,
PKS\,1144-379, or PKS\,1921-293) to enable calibration of the interferometer and
antenna pointing at Parkes and Tidbinbilla. 

We recorded two 16 MHz bandpasses ($\sim 215\kms$) at each station with an S2 VLBI
terminal (Table\,1). The S2 correlator at the Australia Telescope (AT)
facility in Marsfield, NSW provided 1024 channels in each bandpass for cross-power
and total-power spectra, following Fourier inversion of the correlation functions. 
The channel spacing was $\sim 0.21$\kms, which provided at least three channels
across the half-power full-width of the narrowest spectral components \citep[e.g.,
see][]{nakai95, greenhill97}.

\section{Calibration and Imaging}

We calibrated the amplitudes, delays, and phases of the data with
standard VLBI techniques, using the AIPS package of the NRAO\footnote{The  National
Radio Astronomy Observatory is operated by Associated Universities, Inc., under
cooperative agreement with the National Science Foundation.} and the module ATLOD
that is maintained separately by the AT.  

We computed a time-series of amplitude calibration factors for each station  by
fitting total-power spectra for the line triplet near
565\kms~(Figure\,\ref{spectra}) to template total-power spectra, constructed from
data recorded at Mopra in 1997 July and Tidbinbilla in 1998 June. 
(We assume the radio astronomical definition of Doppler shift and heliocentric
reference frame.)   This calibration corrected for variations in antenna gain and
atmospheric opacity.  The template scans were calibrated with nominal system
equivalent flux densities (SEFD) of 900 and 150~Jy, respectively, which provided a
flux density scale accurate to $\sim30\%$, given the elevations of the template
scans and overall good weather conditions.    The weak line strength and irregular
spectral baselines at several stations in 1997 June precluded template
fitting.  Instead, amplitude calibrations were adjusted to maintain constant measured
amplitude for the lines near $\sim 565$\kms~in the {\it cross-power} spectra, which
is appropriate for relatively point-like emission.  A total-power spectrum from
Mopra, calibrated with the nominal SEFD, provided the absolute amplitude
calibration.  

We note that the quality of these amplitude calibrations was limited by the effects
of interstellar scintillation, which in one instance has caused  a doubling in line
strength on time scales as short at 10 minutes \citep{greenhill97}.  However, for the
1998 June and 1997 July epochs the effects of scintillation appear to have been
relatively small.  The total-power spectra constructed for the template fitting show
variations in line strength of $\la 30\%$ over time scales on the order of 30
minutes.  Because we could not construct reliable templates for the 1997 June data,
that epoch could have been more significantly affected by scintillation.  This would
reduce the dynamic range of images but not affect the centroid positions of
individual emission features. 

We used the calibrator scans to estimate delay, fringe rate, and fringe phase
residuals induced by errors in the {\it a priori} station clocks and positions, and 
by electronics and the troposphere.  We interpolated the delays and rates and
applied them to scans of the maser. For the 1998 data, the calibration was accurate
to $\sim2$ ns and $\sim2$ mHz in delay and rate, respectively.  For the 1997 data,
the accuracy for Mopra was worse, $\sim5$ ns and $\sim 5$ mHz, because the maser
clock was inadequately shielded against the  (pointing-dependent) magnetic signature
of the telescope structure.   We also used observations of the calibrators
PKS\,1424-418 and PKS\,1921-293 to estimate the time averaged (complex) bandpass
response of each station.  Overall, residual errors in calibration tended chiefly to
scatter power in the synthesis images and reduce the dynamic range.  Induced errors
in the positions of emission features were ultimately much less than the size
interferometer beam and did not affect the modeling.

The {\it a priori} position of the maser was another potential source of delay
calibration error.   \citet{gw82} found that the emission lay within 10 to 15$''$ of
the optical center for the nucleus \citep{freeman77}, which leaves a very large
uncertainty.  We analyzed the time variation of fringe rate for the 565\kms~line
triplet observed in 1997 July to improve the astrometric position.   A
$\chi$-squared minimization for a point-source model yielded a new position estimate,
$\alpha_{2000}=14^h13^m09\rlap{.}^s95\pm0\rlap{}^s.02$,
$\delta_{2000}=-65^\circ20'21\rlap{.}''2\pm0\rlap{.}''1$, where the quoted
uncertainties include random and systematic components.  This position is offset
from the {\it a priori} position by $+1\rlap{.}''5$ and $+0\rlap{.}''1$ in right
ascension and declination, respectively.  We corrected the 1997 July data and used
the improved position estimate in correlation of the 1997 June and 1998 June epochs,
after which the residual fringe rates were $\sim2$ mHz (peak-to-peak) on even the
longest baselines. This is consistent with the remaining 
$0\rlap{.}''1$ uncertainty.

To remove the effects of atmospheric path-length fluctuations at each epoch, we
self-calibrated the emission within a few\kms~of 565\kms~and applied the results
to each spectral channel. We adjusted the weighting of data in the {\it(u,v)}-plane
and the frequency-averaging of visibility data for each epoch, so that we could
achieve a mean angular resolution of $\sim 2.5$~milliarcseconds (mas) and good
signal-to-noise ratios ($S/N$).  (We convolved the data with a 1.3\kms~wide boxcar, 
sampling every 0.65\kms~for 1997 June, a 0.87\kms~boxcar with sampling every
0.44\kms~for 1997 July, and a 0.44\kms~boxcar with sampling every 0.44\kms~for 1998
June.)  The noise in the images ($1\sigma$) was 0.025 - 0.045~Jy, depending on the
epoch and channel; the 1998 June observations were the most sensitive and best
calibrated, with the broadest bandwidth (Table\,1).  To test the completeness of
our census of emission detected in 1998 June, we also constructed images with natural
weighting and achieved a sensitivity of 15~mJy over 0.44\kms.  In these images, we
detected two new emission clumps at the red and blue extrema of the measured
spectrum. In the end, we detected emission between 179 and 699\kms
(Figure\,\ref{spectra}).

The synthesized beamwidths were $3.7\times 2.0$~mas at PA
$=77^\circ$ (1997 June), $3.7\times 1.4$ mas at PA $=80^\circ$ (1997 July),
$3.6\times 1.9$ at PA$=84^\circ$ (1998 June 27), and  $4.0\times 1.4$ mas at
PA=$-79^\circ$ (1998 June 28).  We fit a 2-D Gaussian model brightness distribution
to each emission ``maser spot'' stronger than $5\sigma$ in each deconvolved image. 
Because the image pixels are
partially correlated, we adopted a position uncertainty for each fitted
Gaussian that was (1) the formal fitting error or (2) the statistical error dictated
by the beamwidth divided by $2\times S/N$ \citep{rm88, fomalont99}, whichever was
larger.  To this uncertainty, we added in quadrature a term accounting for the
effect of possible error in the absolute position of the 565\kms~emission
($0\rlap{.}''1$). This term contributed 35$\mu$as per 100\kms~of velocity offset
from $565\kms$.  For 1998 June and 1997 July, the spreads in spot positions for
individual  clumps of emission are consistent with the computed total position
uncertainties.  The larger spreads observed in 1997 June are suggestive of an
unmodeled systematic component of position uncertainty, probably stemming from the
lower quality calibration and limited coverage of the {\it(u,v)}-plane
for that epoch.

\section{Results and Interpretation}

At each epoch, the sky distribution of maser emission appears to comprise two
populations (Figure\,\ref{4panel}) that we will argue sample a roughly edge-on warped
accretion disk as well as an outflow of material from the vicinity of the
central engine. There is a thin, densely sampled, {\cmss S}-shaped locus 
that comprises redshifted emission to the west, and blueshifted
emission to the east.  The arms of the {\cmss S} exhibit velocity gradients along
their lengths with increasingly large Doppler shifts (relative to the systemic
velocity) toward the center.  Most of the remainder of the maser emission
comprises knots that lie outside the {\cmss S} and away from the putative disk. 
However, even when relatively close to the {\cmss S}, the line-of-sight velocity
of the non-disk maser emission is offset  by tens of \kms.   We propose that this
non-disk emission originates in a wide-angle nuclear outflow.

The velocities of the most red and blueshifted maser spots, 179\kms~and
699\kms~(Figure\,\ref{spectra}), symmetrically bracket the nominal systemic
velocity of Circinus, $439\pm2$\kms \citep{freeman77,curran98}.  The overall spectrum
and sky distribution changed from epoch to epoch, with emission features coming and
going, probably because of natural variation in the intensities of individual
emission features, as well as interstellar scintillation.  To obtain the most
complete assessment of the extent of dense molecular structure that underlies the
maser emission in Circinus, we superposed the maps for two well separated epochs,
1997 July and 1998 June (Figure\,\ref{combomap}).  To register the two maps, we
aligned the positions of the common emission peak at 565.2\kms.  A comparison of
relative positions for other features that are present in both maps suggests the
registration uncertainty is $\la 0.15$~mas.

\section{Discussion}

\subsection{The warped disk}

We interpret the distribution of maser emission in Circinus in the context of the
warped accretion disk in NGC\,4258, which has been well modeled
\citep[e.g.,][]{hgm96}. In general, maser emission in warped disks is beamed
toward the observer from regions where the gradient in line-of-sight velocity along
the line of sight is close to zero and the disk is approximately tangent to the line
of sight.  This corresponds to two loci, the ``midline,'' where the orbital velocity
is parallel to the line of sight, and an arc along the near side of the disk, where
the orbital velocity is nearly perpendicular to the line of sight.  The former is
responsible for high-velocity emission that traces the  rotation curve of the disk,
and the latter is associated with low-velocity emission.  The detailed spatial and
velocity distribution of observable high- and low-velocity emission is also affected
by the geometry of the warp, which (1) determines the solid angle into which
emission is beamed, (2) shadows parts of the disk from X-ray emission by the
central engine, which otherwise heats gas and probably supports maser emission
\citep{neufeld94, nm95, collison95, ww97}, and (3) determines how the disk might
overlie and thereby amplify background nonthermal continuum emission, such as from a
jet \citep[e.g.,][]{herrnstein97}.

A warped disk model for Circinus (see Table\,3) receives its strongest support
from the following: (1) the elongated, antisymmetric angular distribution of
maser spots that forms the shallow  {\cmss S} in Figure\,\ref{combomap}, (2) the
symmetric bracketing of the nominal systemic velocity by the most highly red and
blueshifted emission, (3) the antisymmetric distribution of Doppler shifts
along the {\cmss S} with velocity declining by roughly $b^{-0.5}$, where $b$
is impact parameter (Figure\,\ref{posvel}), and (4) the orientation of the
inner disk (as seen in projection), more or less perpendicular to the axis of
the known radio lobes \citep{elmouttie98} and ionization cone
\citep{marconi94, vb97, wilson00}.

To establish which maser spots arise in the disk, we fit a second order polynomial
to the sky position of the midline, which is identifiable because that is where the
greatest Doppler shifts are observed (Figure\,\ref{combomap}), and we calculated the
angular offset between each maser spot and the midline.  We classified emission that
is offset by $<1$ mas (about one half beamwidth, north-south, in 1998 June) as disk
emission (Figure\,\ref{disk.vs.wind.map}). About 43\% of the integrated emission we
detected in 1998 June arises in the disk, and on average, maser spots in and out of
the disk are comparably bright. The non-disk emission alone occupies the center
$\sim 260$\kms~of the spectrum, while the disk emission lies on either side in
velocity (Figure\,\ref{disk.vs.wind.spectra}).  There is some overlap in velocity,
the extent of which depends on the chosen cutoff to the angular offset from the
midline (i.e., here, 1 mas).  This overlap may be a consequence of resolved disk
thickness, azimuthal structure and a not quite edge-on orientation, or interaction
between the disk and the outflow whereby the fastest spots (in projection)
within the outflow lie closest to the disk.

The position of the dynamical center of the disk, its inner (observed) radius,
and peak rotation speed may be estimated from the innermost red and blueshifted
masers, for which we assume a common edge-on orbit.  The peak observed rotation
speed, $V_{max}$, is 260\kms~in an orbit at a position angle of $29^\circ$
with radius, $r_{\rm in}$, of $5.3\pm 0.1$~mas ($0.11\pm0.02$~pc, including
the uncertainty in distance).  The inferred disk center lies at 17.6 and
13.1~mas east and north of the origin in the maps in Figures\,\ref{4panel} and
\ref{combomap}. Taking into account only gravity and assuming circular motion,
we infer an enclosed mass, $M(r<r_{in})$, of $1.7\pm 0.3\times10^6$~M$_\odot$
or mean mass density of $3.2\pm0.9\times10^8$~M$_\odot$\,pc$^{-3}$.  We note
that because the disk does not display identifiable low-velocity emission on
the near side, it is difficult to estimate inclination directly (cf.
NGC\,4258; Miyoshi\etal 1995). Formally, the mass and density estimates depend
on the inclination and are lower limits. However, we argue that the disk must
be close to edge-on because the masers
cluster tightly around a sharply defined disk midline.

The orientation of the innermost orbit on the disk is close to perpendicular to the
flow axes of the known bipolar radio lobes \citep{elmouttie98} and wide-angle CO
outflow cone \citep{curran99}, the rotation axes of the H\,I galactic disk
\citep{freeman77} and the circumnuclear CO disk of radius 140 to 600~pc
\citep{curran98}, and the principle axis of the most prominent [O\,III] filament in
the ionization cone, for which the position angle is $\sim -50^\circ$ \citep{vb97}. 
However, the assumed $90^\circ$ inclination of the disk at the inner radius is
somewhat offset from the $65^\circ\pm2^\circ$ inclination of the galactic disk
\citep{freeman77}, the $78^\circ\pm1^\circ$ inclination of the CO disk
\citep{curran98}, and the $\sim 65^\circ$ inclination inferred by \citet{wilson00}
for a 40~pc-radius ring of H\,II regions centered on the central engine.   Some of
these alignments are sensible, such as the position angles of the inner accretion
disk and the radio lobe axis.   However, others may include an element of
coincidence, such as the alignment between the  circumnuclear CO disk and the
accretion disk, for which agreement with the inner orbital axis is better than
agreement with the outer orbital axis (Table\,3).  Overall, we suggest that the
orientation of the accretion disk should be only weakly coupled via gravity to the
surrounding large-scale dynamical structures, because the central engine contributes
only $\sim 0.5\%$ of the $3.2\times 10^8$~M$_\odot$ inside 140~pc and has a sphere
of influence less than a few parsecs in radius \citep{curran98}.

The inferred peak rotation velocity as a function of impact parameter, $b$, measured
from the estimated dynamical center, declines approximately as $b^{-0.5}$
(Figure\,\ref{posvel}), which is suggestive of a relatively low-mass disk, as is 
observed in NGC\,4258 \citep{miyoshi95}.  However, a disk that is both massive and
warped in inclination angle could have a similarly steep rotation curve.
Specifically, if a disk is edge-on at radius $r_\circ$ but inclined by 
$\delta i$ at a radius $r+\delta r$, then it will display the same apparent rotation
speed as a flat, edge-on, Keplerian disk if the fractional disk mass $M_{\rm
disk}/M$ is $\tan^2(\delta i)$, where $M$ is the encircled mass at radius $r_\circ$.
In order to place a limit on the disk mass for Circinus, we assume that the disk is
warped both in  position and inclination angles, and that the warps are of the same
order of magnitude, $\sim 27^\circ$ between 0.11 and 0.40~pc (see Table\,3).  Hence,
our data are consistent with a fractional mass of 26\%, or $\sim 4\times
10^5$~M$_\odot$.  In the absence of an actual measurement of the inclination warp,
we adopt this mass estimate as an upper limit.  

The relatively high particle densities inferred from the presence of H$_2$O  maser
emission --- $10^8$ to $10^{10}$~cm$^{-3}$ for a H$_2$O fractional abundance of
$10^{-5}$ to $10^{-4}$ \citep{elitzurbook} --- are also consistent with a relatively
massive disk.  For a uniform particle density and a planar disk, the disk mass
enclosed inside radius $r$ is $\sim 1.4\times10^6 n_{10}\gamma r_{-1}^3$ M$_\odot$,
where $r_{-1}$ is the radius in units of 0.1~pc, $n_{10}$ is density in units of
$10^{10}$~cm$^{-3}$, and $\gamma$ is the height of the disk as a fraction of its
radius.  Hydrostatic equilibrium dictates a lower limit on $\gamma$ that is on the
order of $c_s/v_{\rm rot}$, where $c_s$ is the sound speed  ($\sim1.6$\kms~for gas
at 400~K).  Our upper limit on disk mass inside 0.40~pc ($\sim 4\times
10^5$~M$_\odot$), is consistent with a $4\times10^9$~cm$^{-3}$ upper limit on
density, for $v_{\rm rot}\sim 140$\kms~and hydrostatic equilibrium, which agrees
with expectations.

Estimates of the Toomre Q-parameter for gas densities sufficiently high to support
maser action ($> 10^8$~cm$^{-3}$) indicate that self-gravity is important and
probably causes clumping of disk material.  We note that  the distributions of disk
masers on the sky (Figure\,\ref{combomap}) and in the position-velocity diagram
(Figure\,\ref{posvel}) exhibit substantial substructure, which may be a signature of
clumping. This clumping  could limit the maser gain paths in the disk and thus be
responsible for the comparable strengths of the disk and non-disk emission.  The
Toomre
$Q$-parameter is $c_s\Omega / \pi G\Sigma$, where $\Omega$ is the angular rotation
speed, and
$\Sigma$ is the mass surface density. Instabilities arise for $Q<1$.  By assuming
the disk thickness is $\eta$ times the hydrostatic limit, we obtain $Q\sim0.6 M_6 /
\eta n_{10} r_{-1}^3$, where
$M_6$ is the central engine mass in units of $10^6$~M$_\odot$.  For
$n=4\times10^9$~cm$^{-3}$, $r=0.40$~pc, and hydrostatic equilibrium ($\eta = 1$), 
$Q\sim0.04$.  The Q-parameter could
be even smaller if the apparent thickness of the disk (which is on the order
of a few tenths of a mas) is real and thereby much greater than the
hydrostatic limit (which is about tens times smaller).  However, it is
difficult to distinguish between measurable disk thickness and nonaxisymmetric
structure \citep[e.g.,][]{mm98}, because the disk is highly inclined  and the
angular resolution of our observations is an order of magnitude larger than
the putative thickness.

\subsection{The clumpy wide-angle outflow}

The association of water maser emission with a bipolar wide-angle wind in an AGN is
qualitatively new and distinct from the reported connection to jet activity in some
galaxies [i.e., NGC\,1068 \citep{gallimore96}, NGC\,1052 \citep{claussen98},
NGC\,3079 \citep{trotter98}, and Mrk\,348 \citep{peck2001}].   Less than 1~pc from a
central engine, outflowing material is often assumed to be photoionized, but away
from the flow axis, gas columns may be large enough ($10^{22\pm1}$~cm$^{-2}$) to
maintain a molecular component with internal pressures ($p/k$) that are $\sim
10^{11}$ to $10^{13}$~K\,cm$^{-3}$ \citep{emmering92, kk94, kke99}.  Maser emission
can be readily stimulated in this gas by X-ray heating \citep{neufeld94,nm95} or
shock heating \citep{ehm89,kn96}, if turbulence and gradients in
line-of-sight velocity permit.

In Circinus, blueshifted ``outflow masers'' lie south and east of the central engine,
and redshifted counterparts lie chiefly to the west (Figures\,\ref{4panel},
\ref{combomap}, \& \ref{disk.vs.wind.map}).  The line-of-sight velocities of these
masers are offset from the systemic velocity by up to $\sim \pm
160$\kms~(Figure\,\ref{disk.vs.wind.spectra}), though we note a few maser
spots that are somewhat more redshifted and that lie almost due  west of the
dynamical center in 1997 July. To start, we infer that this emission originates
outside the disk because (1) it lies up to
$\sim 1$~pc from the disk center, while the disk midline can be traced only between
radii of 0.11 to $\sim 0.40$~pc, (2) many of the non-disk masers lie 
close to the disk rotation axis, as seen in projection,
(3) the line-of-sight velocities of
the non-disk masers that instead lie close to the disk differ by tens of \kms~from
the velocities of adjacent disk material, and (4) the non-disk masers display no
systematic gradient in line-of-sight velocity as a function of position angle
relative to the disk center (Figure\,\ref{mapzoom}), as would be expected were they
to lie in a disk sufficiently warped to include them.  Having excluded a disk
origin, we infer that the masers in question are associated with outflow because it
is unlikely that material close to the axis of an accretion disk, and also within
1~pc of a supermassive blackhole would be infalling.  (We note that it is also
unlikely that the non-disk masers are associated with  a relaxed central cluster of mass
losing stars, principally because the distribution of masers is not at all centered
on the inferred position of the central engine.)

We adopt a  bipolar, cone-like model geometry to describe the outflow
(Figure\,\ref{model}).  Emission with different Doppler shifts is well mixed on the
sky (e.g., Figure\,\ref{mapzoom}). This is suggestive of relatively short gain paths
peppering the outflow volume, where emitting regions that move along very different
trajectories lie along proximate lines of sight (i.e., the masers probably do not
just lie along the limbs of the outflow).  Within this volume, the masers presumably
do lie toward the front because of the opacity of the photoionized gas that probably
lies between the maser-emitting regions (see below).

Although a compelling general case may be made for outflow, we cannot explain the
observed angular distributions and Doppler shifts of the outflow masers in detail. 
The chief hurdles are (1) the prevalence of blueshifted emission to the south and
east of the disk center and redshifted emission on the reverse side, and (2) the
difference in angular distributions on opposite sides of the accretion disk, where
some areas are devoid of visible maser emission while their counterparts (mirrored
on the other side of the disk) are not (Figure\,\ref{combomap}).  The blue-red
asymmetry in the outflow could indicate that the flow is inclined to the line of
sight, despite the putative edge-on orientation of the accretion disk.  However, it
is difficult to test this hypothesis because the redshifted outflow is poorly
sampled on the sky and almost entirely defined by maser emission at low latitudes
(i.e., in close proximity to the disk).  Gas dynamic or magnetic coupling between
outflowing and rotating disk material could be responsible for the observed Doppler
shifts at low latitudes, and in position-velocity space, for some  portion of the
structure we observe near (the representation of) the disk midline (see
Figure\,\ref{posvel}).  Regarding the second hurdle, the difference in maser
distribution above and below the disk could be a consequence of differential
extinction.  However, if the axis of the outflow is close to the plane of the sky,
then this effect would be small.  Alternatively, the angular distributions could
depend chiefly on where velocity coherent maser gain paths happen to exist.  In
general, the longest paths are probably associated with limb-brightening and motions
that are closer to the plane of the sky than to the line of sight
\citep[e.g.,][]{ehm92}.  This observation may answer the question, why do we not see
highly blueshifted outflow material close to the line of sight toward the central
engine. However, it is not clear whether it can explain the absence of a broad
distribution on the sky of redshifted outflow masers.  

Because water masers require high gas densities, their presence broadly distributed
in the outflow argues that it must be inhomogenous or ``clumpy.''  Were the outflow
smooth, the associated mass loss would be prohibitively large, $\sim 6\times10^3
r_{-1}^2 n_{10} v_2$~M$_\odot$\,yr$^{-1}$, where  $v_2$ is velocity in units of
100\kms. For a 100\kms~flow that is $10^9$~cm$^{-3}$ at a radius of 0.1~pc, the flow
would consume a mass equivalent to the central engine and accretion disk in only a
few $\times 10^3$~years.  However, the appearance of the flow, as traced by maser
emission, is at best sparse and irregular.  If this is representative and not merely
a consequence of anisotropic beaming of maser radiation, then the mass loss would
be orders of magnitude smaller.  For instance, if the rate of loss balances the rate
of accretion, then the mean density of the outflow at 0.1~pc radius is $\sim
10^4$~cm$^{-3}$,  where we express the accretion rate as $1.7\times10^{-4} L_{43}
\epsilon^{-1}$, and where $L_{43}$ is the bolometric luminosity in units of
$10^{43}$ ergs\,s$^{-1}$ and $\epsilon$ is the accretion efficiency of the central
engine, probably $\sim 0.1$.  The apparent mean clump-interclump density contrast
is  on the order of $10^5$, and the shielding column in the interclump medium is on
the order of $10^{22}$~cm$^{-2}$ for a path length comparable to the disk radius, as
required  \citep{kke99}. (We note that the flow could instead contain pockets of
molecular gas with a range of sizes and densities that span orders of
magnitude, only some of which would be conducive to observable maser action.  This
would affect the estimates of density contrast and shielding column, but it
would not reduce the imperative that the flow be inhomogeneous.)

\subsection{Collimation by the warped disk}
 
The angular distribution of the outflow masers is consistent with the
hypothesis that excitation depends on a direct line of sight to the central
engine.  In the shadow of the disk (i.e., in regions where this path is blocked)
no maser emission was detected (Figure\,\ref{model}). We suggest that this
shadowing is a physical blockage of both radiation and mechanical energy through
which {\it the warped accretion disk collimates the nuclear outflow.}  Channeling on
such small scales is consistent with the inference by \citet{wilson00} based on HST
imaging, that the Circinus ionization cone is collimated within 2~pc of the central
engine.  In the proposed model, on each side of the maser disk, one radial edge of
the outflow is determined by the position angle of the disk at the inner radius, and
the other edge is fixed by the concave surface of the disk warp along which the
outflow brushes, up to the outer radius of the disk. 

We estimate the outer radius to be $\sim 0.40$~pc based on the close packing of disk
masers along the midline up to this radius, the angular distribution of outflow
masers at larger radii, and the extent of the ionization cone. First, the midline of
the warped disk is traced in the maps (Figure\,\ref{combomap}) and the
position-velocity diagram (Figure\,\ref{posvel}) by a relatively continuous locus of
maser spots up to a radius of $\sim 0.40$~pc.  Beyond 0.40~pc, an extrapolation of
the midline does pass close to some other masers, but they do not lie on the
Keplerian rotation curve.  Second, the only way in which the southwesternmost maser
clump near position (-16, -8)~mas in Figure\,\ref{combomap} can achieve a direct
line of sight to the central engine is if the disk is truncated at $\sim 0.40$~pc.
Third, the southern edge of the ionization cone, observed in [O\,III] light
\citep{marconi94, vb97, wilson00}, lies at a position angle of $\sim -120^\circ$,
which is close to the position angle of the disk midline at a radius of $\sim
0.40$~pc.  Were the warped disk to extend substantially beyond this proposed outer
radius, the disk would block a larger solid angle of radiation and create a
different southern limit for the ionization cone.  We speculate that  an outer
radius of $\sim 0.40$\,pc may also be associated with the presence of a maser spot
cluster surrounding the outer tip of the redshifted midline, forming a backward ``C''
visible near the origin in Figures\,\ref{combomap} and
\ref{disk.vs.wind.map}. These masers move with sub-Keplerian velocities and  could
correspond to material lost from the outer disk where it is roughened by interaction
with outflow and radiation.

The putative outer radius of the accretion disk constrains the longevity of activity
in Circinus at the observed luminosity, $4\times10^{43}$~ergs\,s$^{-1}$.  The
wide-angle outflow restricts the solid angle (as seen from the central engine) over
which material can fall in and replenish the accretion disk.  The outflow blocks
80\% to 90\% of the surroundings.  If the supply of new material is thereby
reduced, then for the accretion rate ($1.7\times10^{-4} L_{43}
\epsilon^{-1}$) that corresponds to an efficiency of 10\%, the accretion disk will
be exhausted in $\sim 6$~Myr for the previously estimated maximum disk mass of
$4\times10^5$ M$_\odot$. 

\subsection{The Origin of the Outflow}

Two possible origins for the molecular outflow are: (1) radial ejection of ionized
broad line clouds that form dust and molecules as they cool, and (2) uplift and
ejection of warm molecular disk material.  The association of masers and
cooled broad line cloud material is largely circumstantial.  It rests on the
observation that the densities and temperatures required for maser emission are
consistent with adiabatic expansion of material with intial densities and
temperatures on the order of $10^{11}$~cm$^{-3}$ and $10^4$~K \citep[e.g.,][and
references therein]{kn99,kg00}, assuming an adiabatic constant of $\sim {5 \over
3}$.  However, it is not clear whether maser action is possible in freely expanding
clouds, where velocity gradients and turbulent motions may be significant, or in the
shocks that may arise within or ahead of the clouds. The observed line-of-sight
outflow velocities, which are smaller than the Keplerian rotation velocity and
thereby the escape velocity at all radii, are a secondary concern. Because
the motions of the observable masers may lie somewhat preferentially toward the plane
of the sky (see Section\,5.2), the too small velocities may be a projection effect. 

In order to describe self-consistently the geometry and dynamics of broad line
regions, \citet{emmering92} modeled a hydromagnetic wind created by the uplift and
ionization of clumps from a cool molecular accretion disk.  The magnetic field,
anchored in the disk, confined the ionized clouds and imparted rotation to the
outflow.  Over time, the clouds moved outwards with increasingly radial trajectories
as the azimuthal field component or cloud coupling to the field weakened \citep[see,
e.g.,][]{bottorff97}.  \citet{kk94} expanded this model by
showing that the hydromagnetic disk wind could be clumpy, dusty, and molecular over
a broad range of latitudes.  They suggested that such a wind might be responsible in
general for the obscuration observed toward type-2 AGN, rather than oft posited
geometrically thick tori.  \citet{kke99} applied the model to the stimulation of
maser emission in the uplifted clumps once they rise from the disk to be
irradiated by the central engine.  (We note that in this picture, the disk
is not directly the source of maser radiation.)

The hydromagnetic wind model is somewhat more attractive than the simple  ejection
model for broad line clouds principally because in the former the clumps are
magnetically confined. They do not need to expand by orders of magnitude and remain
coherent to achieve temperatures and densities conducive to maser action.  However,
we note that in the limit where outflowing material moves largely radially well
after uplift, the models are similar.  The simple ejection model as stated finesses
the origin of the broad line clouds, but if the clouds originate in an accretion
disk, then the models differ chiefly in whether the disk material remains neutral
after it is removed from the disk. 

However, the hydromagnetic wind model, as applied by \citet{kke99} makes three
predictions that are not easily reconciled with the observations.  First, the masers
should lie away from the rotation axis of the disk.  In contrast, we observe outflow
masers that are apparently quite close to the rotation axis.  If this material is
removed from the observed accretion disk (between radii of $\sim 0.1$ and $\sim
0.4$~pc), then it would have to travel almost vertically as it is uplifted. Second,
masers in a hydromagnetic wind should exhibit a sub-Keplerian rotation curve (i.e.,
in a position-velocity diagram the masers would at least in part lie above the disk
rotation curve).  In fact, the signature of rotation is weak if present at all, and
for any given radius, the Doppler shifts of outflow material are smaller than for
the disk.  Third, the wind should support maser emission close to the loci where
dust sublimation occurs, resulting in a more or less flattened {\cmss X}-pattern on
the sky for relatively edge-on orientations.   However, the actual distribution of
masers is dichotomous, featuring a sharply defined, thin, {\cmss S}-shaped disk {\it
and} a broad outflow component.

We propose a modified wind model.  First, the outflow begins well inside the
observed inner radius of the disk and thereby can reach high latitudes at
radii of 0.1 to 0.4~pc more readily. Very high velocity maser emission, offset
up to $\sim 400$\kms~from the systemic velocity, has recently been discovered in
Circinus \citep{samba1}.  If the emission originates in the accretion disk, it
provides evidence for a resevoir of molecular material $\sim 0.03$~pc from the
central engine.   Second, the uplifted material moves by and large radially, perhaps
because it decouples quickly from the magnetic field threading the accretion disk. 
Third, the outflow is sparsely populated with clumps of material, so that the
warped accretion disk at larger radii is not significantly shadowed and can emit
maser radiation, as is observed.

\section{Conclusions and Comment}

We find that the H$_2$O maser emission in the Circinus Galaxy traces a warped,
edge-on accretion disk at radii between $0.11\pm0.02$~pc and $\sim 0.40$~pc. The peak
observed disk rotation velocity is $\sim 260$\kms, and the encircled mass is
$1.7\pm0.3\times10^6\msun$.  The implied  ratio of bolometric luminosity to Eddington
luminosity is $\sim 0.2$.  The rotation curve is nearly Keplerian and the disk mass
is probably $<24\%$ of the central mass.  Given the (low) rotation speed of the disk,
this is substantial enough to cause clumping due to self-gravity, which may explain
observed substructure in the disk.  The inner radius of the maser disk is comparable
to the well resolved maser disk in NGC\,4258 (0.14~pc), even though the X-ray
luminosity of Circinus is $\sim 10$ times larger \citep{makishima,reynolds00}.  If
the recently reported very high velocity emission in Circinus originates in the disk
\citep{samba1}, then the paradox is more striking still. However, it is possible the
inner radii of maser emission are not dictated by dust sublimation. The discrepancy
may be due to the geometries of the warped disks.  For example, in NGC\,4258 maser
emission at particularly small radii may not be directed toward us. 

A second population of masers in the Circinus AGN traces a wide-angle outflow up to
$\sim 1$\,pc from the central engine with line-of-sight velocities of up to $\sim 
\pm160$\kms~with respect to the systemic velocity.  These masers are the first direct
evidence (i.e., provided by imaging) of dusty, high-density, molecular material in a
nuclear outflow, at such small scales.  The mechanism that drives the wind is
uncertain but may be hydromagnetic.  The wind is observed in the regions around the
disk that are not shadowed by the warp, from which we suggest that the disk channels
the outflow and occults radiation from the central engine.  The position angles of
the edges of the outflow correspond well to those of the outflow and ionization cone
observed on scales of hundreds of parsecs \citep{vb97, curran99}.   Here too,
Circinus may provide the first direct evidence for collimation of an AGN outflow and
illumination pattern by a warped accretion disk.  In the classical paradigm for AGN,
a geometrically thick tori has been cast in this role, even though it has been
difficult to explain how such structures can be supported vertically
\citep[e.g.,][]{kb88}.  For Circinus and perhaps other type-2 AGN, sub-parsec,
warped accretion disks are a simpler explanation of a breadth of observations.  

It is an open question whether thick tori are required at all to explain the X-ray
obscuring columns observed in type-2 AGN.  Warped disks of the type found in
NGC\,4258 and Circinus could easily supply the needed column densities, but they are
relatively flat.  These two disks subtend on the order of 10 to 20\% of the sky as
seen from their respective central engines and do not readily explain the high
proportion of obscured AGN out of the total \citep[e.g.,][]{mg90}.  As suggested by
\citet{kk94}, dusty hydromagnetic winds may be a general source of obscuration. 
Alternatively, structures on the small spatial scales of broad line regions
\citep{risaliti} or the $>100$~pc scales of warped CO gas disks \citep{schinnerer}
and other dusty galactic structures \citep{malkan} may be important.

The Circinus H$_2$O maser is the third one for which detailed VLBI study has
detected disk structure and a declining rotation curve.
As has been demonstrated for the NGC\,4258 and NGC\,1068 masers, the
Circinus maser makes possible unusually detailed study of the structure and dynamics
of a region $<1$~pc from what is most probably a supermassive black hole. We
anticipate that higher angular resolution VLBI observations covering the recently
recognized, full $\sim 900$\kms~breadth of the maser will improve our understanding
of the structure of the $\sim 0.1$~pc radius warped accretion disk and the mechanism
that drives the wide-angle outflow apparently emanating from still closer to the
black hole.

\acknowledgements

We thank W. Wilson and R. Ferris for their support of the S2 VLBI
correlator, and H. May and S. Amy for network and software support. We acknowledge
the helpful comments of G. Ball, M. Elitzur, A. Konigl, and an anonymous referee.  We
are grateful for the dedication of the staff at the LBA antennas.  We acknowledge
the use of the NASA/IPAC Extragalactic Database (NED), which is operated by the Jet
Propulsion Laboratory, California Institute of Technology, under contract with the
National Aeronautics and Space Administration.  

\newpage 

\begin{deluxetable}{lllll}
\tablecaption{Journal of Long Baseline Array Observations}
\label{lba.journal}
\tablehead{
\colhead{Date} & 
\colhead{Array\tablenotemark{(1)}} & 
\multicolumn{2}{c}{Bandpasses\tablenotemark{(2)} } & 
\colhead{Notes} \\ 
\cline{3-4} 
\colhead{} & 
\colhead{} & 
\colhead{(\kms)} &
\colhead{(\kms)} & 
\colhead{} }

\startdata

1997 June 24 & Pk-Mo-Ho-Td    & 233--449 & 454--670 & No fringes at Td for
80\% of track.\\ 

1997 July 25 & Mo-Ho-Td    & 234--450 & 455--671 & No fringes at Pk. \\ 

1998 June 27 & Pk-Mo-Ho-Td & 313--529 & 488--704 & \nodata \\ 

1998 June 28 & Pk-Mo-Ho-Td & 151--367 & 488--704 & \nodata \\

\enddata

\tablenotetext{(1)}{Pk--Parkes, NSW (Parkes); Td--Canberra, ACT (Tidbinbilla);
Mo--Coonabarabran, NSW (Mopra); Ho--Mt. Pleasant, Tasmania (Hobart).}

\tablenotetext{(2)}{Heliocentric velocities with respect to the radio definition
of Doppler shift.  Optical velocities are up to 1.6\kms~larger over
the range of velocities observed.}

\end{deluxetable}

\begin{deluxetable}{llccl}
\label{stations}
\tablecaption{The LBA at $\lambda 1.3$cm (c. 1997/1998)}
\tablehead{
\colhead{Telescope} & 
\colhead{Mount}     & 
\colhead{Diameter}  & 
\colhead{SEFD\tablenotemark{(1)}} & 
\colhead{Receiver}  \\
\colhead{}    &
\colhead{}    &  
\colhead{(m)} & 
\colhead{(Jy)}&
\colhead{}          }      

\startdata
Parkes &Alt-Az& 64       & 700  & cooled HEMT; prime focus \\  

Mopra  &Alt-Az& 22 & 900 & cooled HEMT; cassegrain focus \\ 

Hobart &X-Y\tablenotemark{(2)}& 26 & 5000   & uncooled HEMT; prime focus\\

Tidbinbilla &Alt-Az& 70     & 150 & cooled HEMT; cassegrain focus \\

\enddata
\tablenotetext{(1)}{Nominal values under good weather conditions and moderate 
elevation. Calibration of values for each observation and the elevation
dependence is discussed in section $\S3$.}
\tablenotetext{(2)}{X--Y mount with a north-south axis of motion.}

\end{deluxetable}

\newpage

\begin{table}[ht]
\centerline{Table 3: The Warped Accretion Disk and Central Engine}
\bigskip

\begin{center}
\begin{tabular}{ll}
\tableline\tableline
 Reference$^{(1)}$ $\alpha_{2000}$& $14^h13^m09\rlap{.}^s95\pm0\rlap{}^s.02$ \\
 ~~~~~~~~~~~~~~~~ $\delta_{2000}$   
             &$-65^\circ20'21\rlap{.}''2\pm0\rlap{.}''1$\\          
 $r_{\rm inner}$                    & $0.11\pm0.02$pc\,\tablenotemark{(2)}\\
 $r_{\rm outer}$                    & $\sim 0.40$ pc\,\tablenotemark{(2)} \\
% $V_{\rm systemic}$                 & 439\kms\,\tablenotemark{(3)}       \\
 $V_{\rm orbital}(r=r_{\rm inner})$ & 260\kms\,\tablenotemark{(4)}        \\
 $M(r<r_{\rm inner})$               & $1.7\pm0.3\times10^6$ M$_\odot$     \\
 $\rho(r<r_{\rm inner})$            & $3.2\pm0.9\times10^8$ M$_\odot$\,pc$^{-3}$\\
 Period $(r=r_{\rm inner})$         & $2600\pm500$ yr                     \\
 $\alpha$ ($V\propto r^\alpha$)     & $\sim -0.5$                         \\
 $PA(r=r_{\rm inner})$              & $29^\circ\pm3^\circ$                \\
 $PA(r=r_{\rm outer})$              & $56^\circ\pm6^\circ$                \\
 $M_{\rm disk}(r_{\rm inner}<r<r_{\rm outer})$&
                             $<4\times10^5$ M$_\odot$\,\tablenotemark{(5)}\\

% $r_{\rm subl}$                    & $\sim 0.06$ pc\,\tablenotemark{(6)} \\

\tableline
\end{tabular}
\tablenotetext{(1)}{Position of the maser emission at 565.2\kms~in 1997 July.
The center of the disk (i.e., the central engine) was offset 17.6~mas east and
13.1~mas north.  The optical nucleus is offset
by $-1\rlap{.}''5$ and $-0\rlap{.}''1$ in right ascension and declination,
respectively \citep{freeman77}.}
\tablenotetext{(2)}{For an assumed distance of 4.2 Mpc \citep{freeman77}.}
\tablenotetext{(3)}{Adopted from \citet{freeman77,curran98}.}
\tablenotetext{(4)}{Assuming the disk is edge-on at the inner radius.}
\tablenotetext{(5)}{Computed on the premise that the moderating effect of a
substantive disk mass on the slope of the rotation curve is balanced by the
steepening effect of a warp in inclination, leaving a Keplerian rotation curve.
The largest probable change in inclination (from $90^\circ$ at $r_{\rm inner}$) is on
the order of the observed change in position angle, $\sim 27^\circ$. See
Section\,5.1.}

%\tablenotetext{(6)}{Approximate dust sublimation radius, after \citep{kke99},
%assuming a dust grain size and spectral energy distribution (SED) similar to those
%of NGC\,1068.}

\end{center}
\end{table}

\newpage
\leftline{\bf Figure Captions}
\smallskip

\figcaption[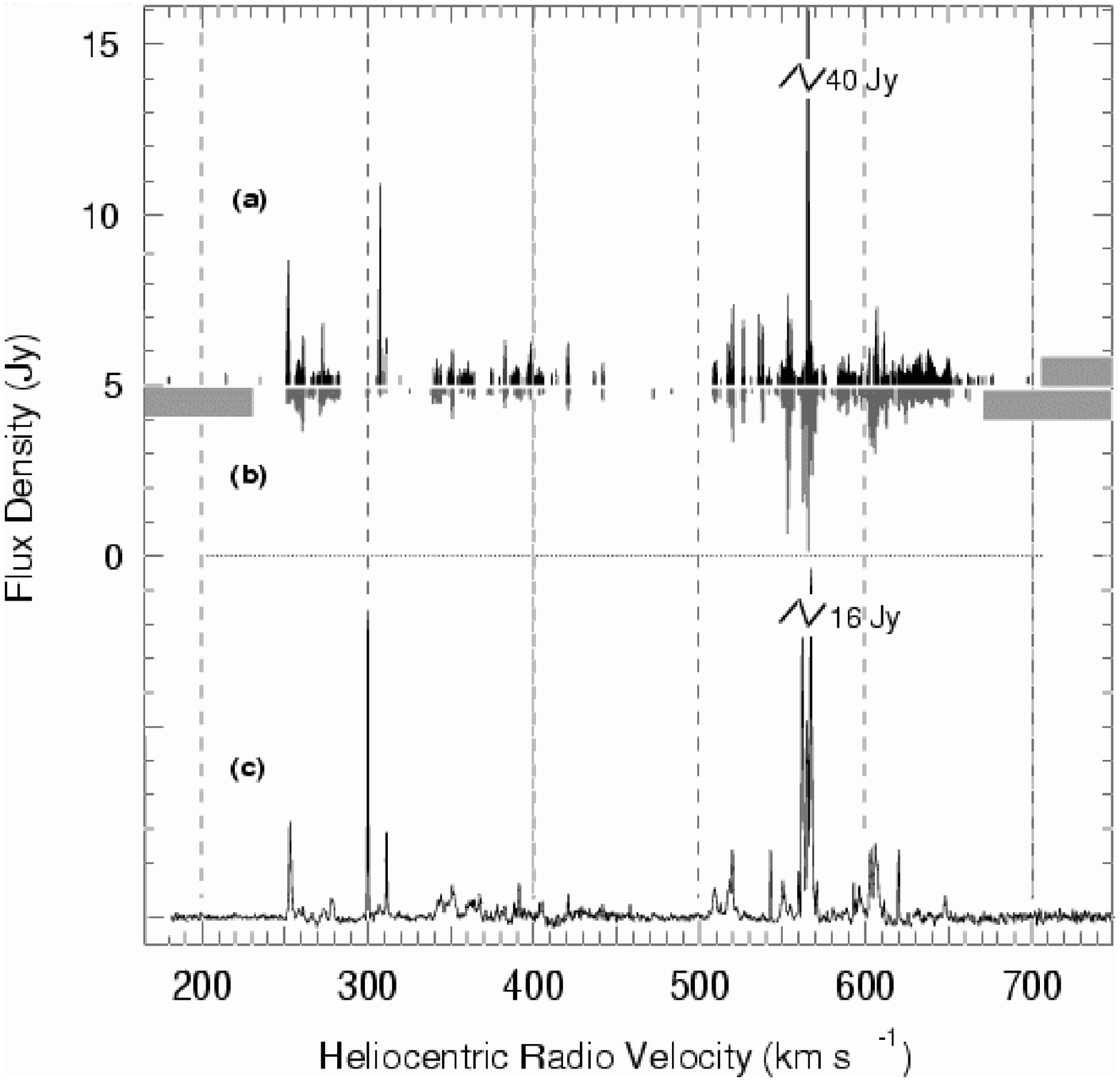]{Maser spectra of total imaged power at two epochs and total
power at one epoch.  {\it (a)} 1998 June total imaged power.  {\it (b)} 1997
July total imaged power (inverted and in gray).  {\it (c)} 1995 total-power
spectrum obtained with the Parkes antenna \citep[taken from][]{greenhill97}.  The
instantaneous bandwidth of the VLBI observations was $400\kms$~in 1997 and
$\sim 600$\kms~in 1998. The gray patches indicate velocities outside that
range.\label{spectra}}

\figcaption[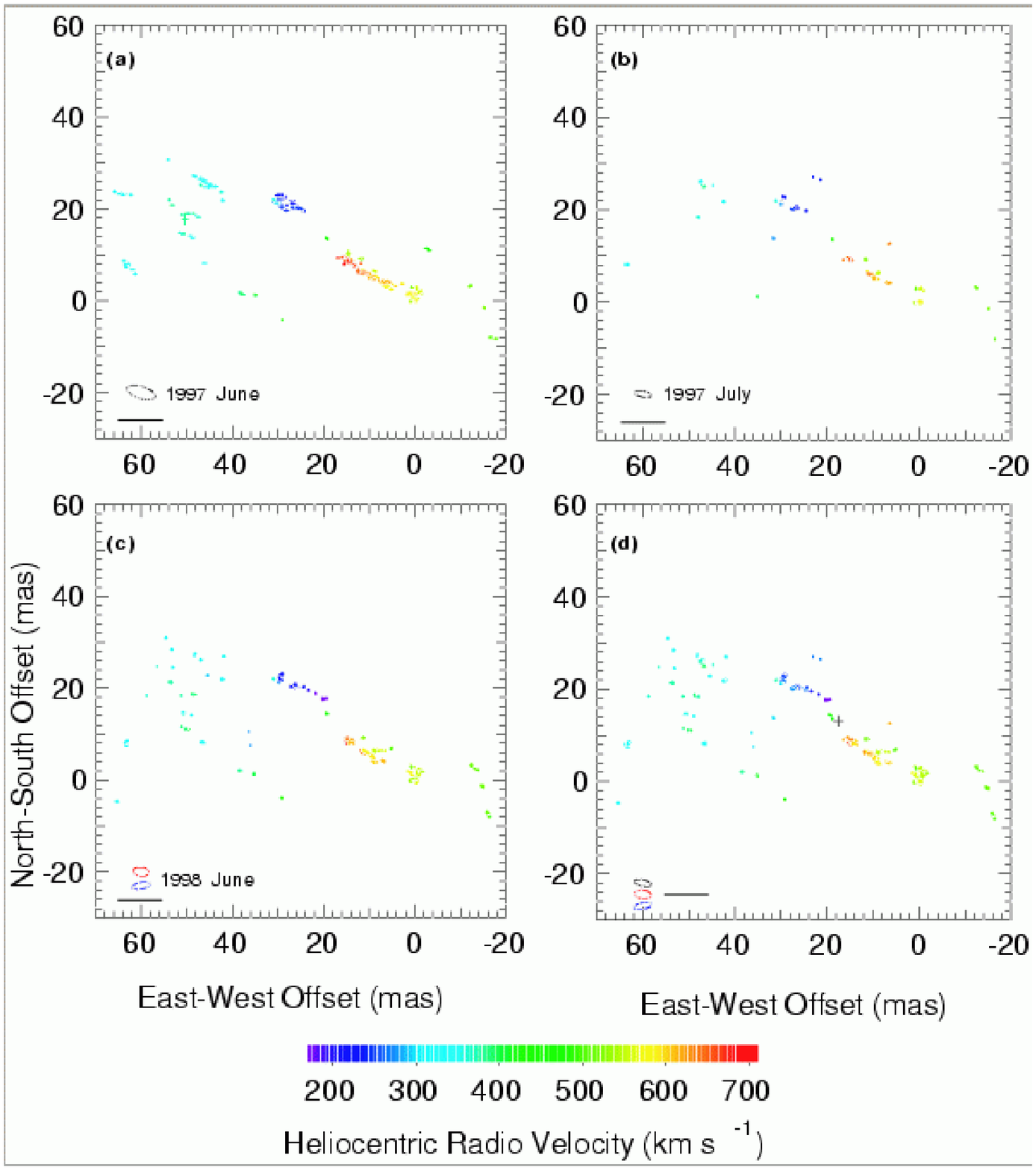]{Sky distributions of H$_2$O maser emission in
Circinus. Each symbol represents a single maser spot (i.e., emission in a single
spectral channel).  Color indicates heliocentric velocity as shown on the scale
below, assuming the radio definition of Doppler shift. The horizontal black scale
bar corresponds to 0.2\,pc at a distance of 4.2\,Mpc. {\it (a)} 1997 June 24. {\it
(b)} 1997 July 25. {\it (c)} 1998 June 27 and 28. {\it (d)} 1997 July and 1998 June
plotted together,  registration is accurate to
$<0.15$ mas, based on the apparent alignment of emission clumps common to both
epochs. The cross marks the estimated dynamical center of the disk (see
Figure\,\ref{combomap}). The 1997 June data are excluded because of  probable
residual calibration errors that are probably responsible for the apparent
elongation of many clumps of emission.  In each map, the origin corresponds to
$\alpha_{2000}=14^h13^m09\rlap{.}^s95\pm0\rlap{.}^s02$,
$\delta_{2000}=-65^\circ20'21\rlap{.}''2\pm0\rlap{.}''1$.  Error bars indicate
uncertainties ($1\sigma$) in position, but in most cases they are smaller than the
plotted symbols. The synthesized beams are shown symbolically in the lower left of
each panel. For the epoch in 1998, there are different beams for the low-velocity and
redshifted high-velocity data (red ellipse) and the blueshifted high-velocity data
(blue ellipse).
\label{4panel}}

\figcaption[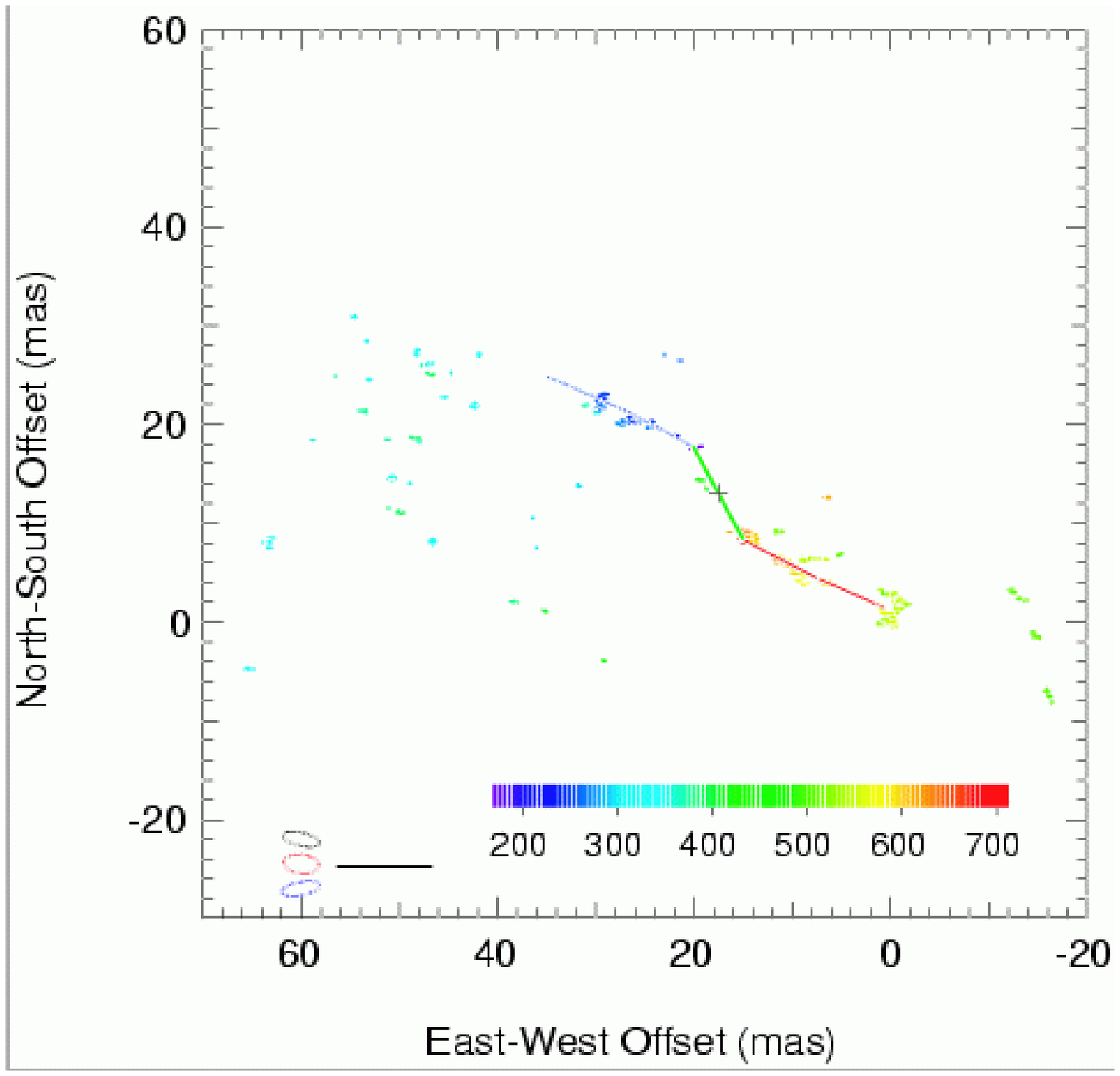]{Superposition of maps for 1997 July and 1998
June. Color indicates velocity, as in Figure\,2. The black 
scale bar corresponds to 0.2\,pc. The blue and red curves indicate the limb or
midline of the putative edge-on warped disk, where the orbital velocity is
parallel to the line of sight. The red curve is a second order polynomial
fitted to the midline.  The blue curve is a reflection of the polynomial about
the center of the disk.  The good agreement between the reflected curve and the
blueshifted midline demonstrates how well the data fits an antisymmetric warp
model. The lengths of the arcs indicate the range of radii over which
redshifted maser spots trace a smoothly declining rotation curve (see
Figure\,\ref{posvel}).  The green line represents the innermost orbit, defined
by the most highly red and blueshifted maser spots. The cross marks the
estimated dynamical center of the disk.
\label{combomap}}

\figcaption[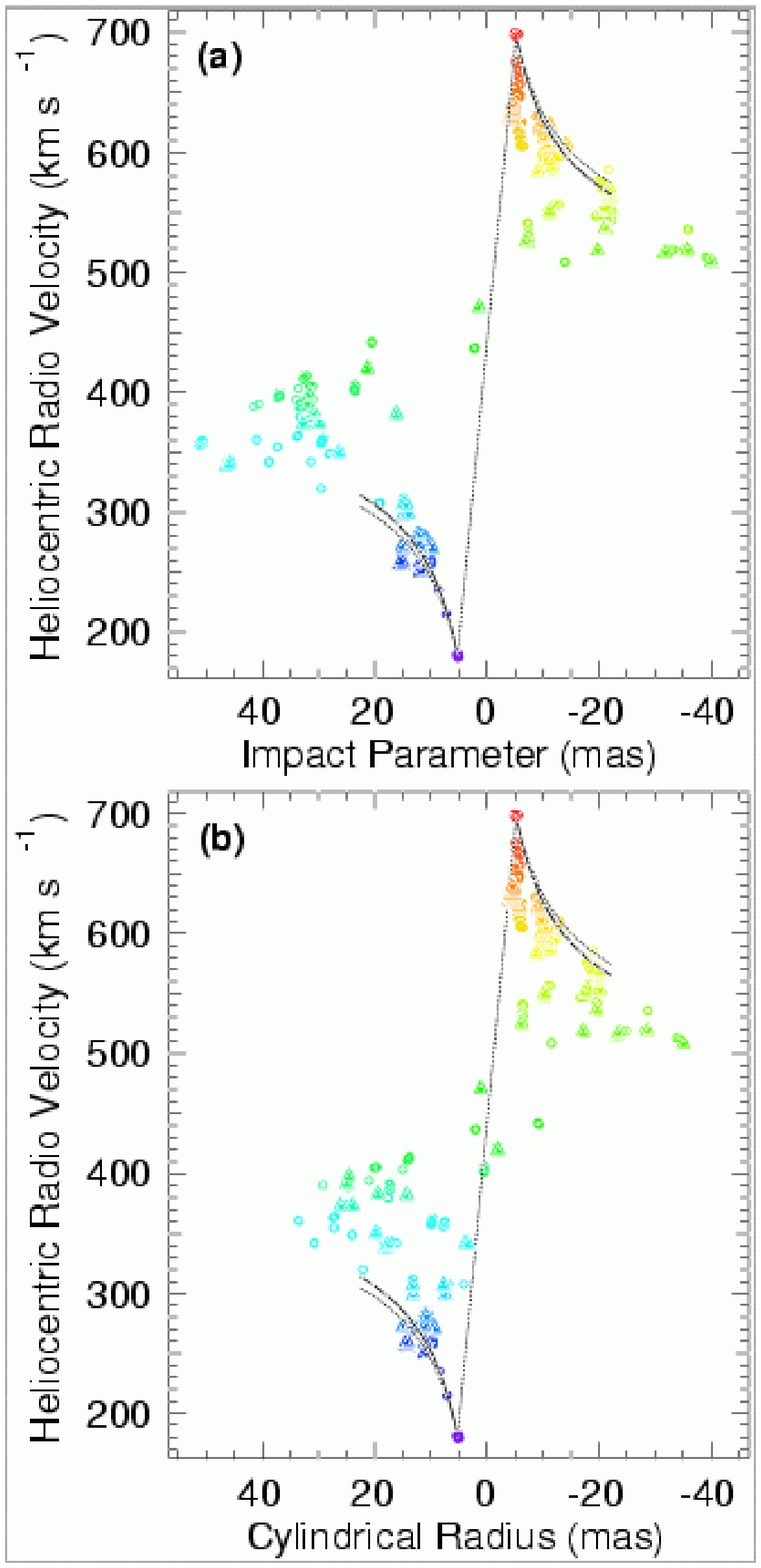]{Position-velocity diagrams combining data for
1997 July {\it (triangles)} and 1998 June {\it (circles)}. {\it (a)} Velocity plotted
against impact parameter measured from the estimated position of the central
engine.  {\it (b)}  Velocity plotted against cylindrical radius measured from the
rotation axis of the innermost orbit of the proposed model disk. 
Both plots are presented because the orbital speed of disk material is a function of
impact parameter while any rotation in the outflow is probably a function of
cylindrical radius \citep[see][]{kke99}.  Solid and dashed rotation curves reflect
$v\propto r^{-0.5}$ and
$v\propto r^{-0.45}$, respectively.  The data points that lie close to these curves
represent masers that arise close to the midline of the disk.  The proposed
wide-angle outflow is represented by the masers that lie inside and away from the
two rotation curves.  The steep diagonal line delineates the locus for the innermost
orbit.  
\label{posvel}}

\figcaption[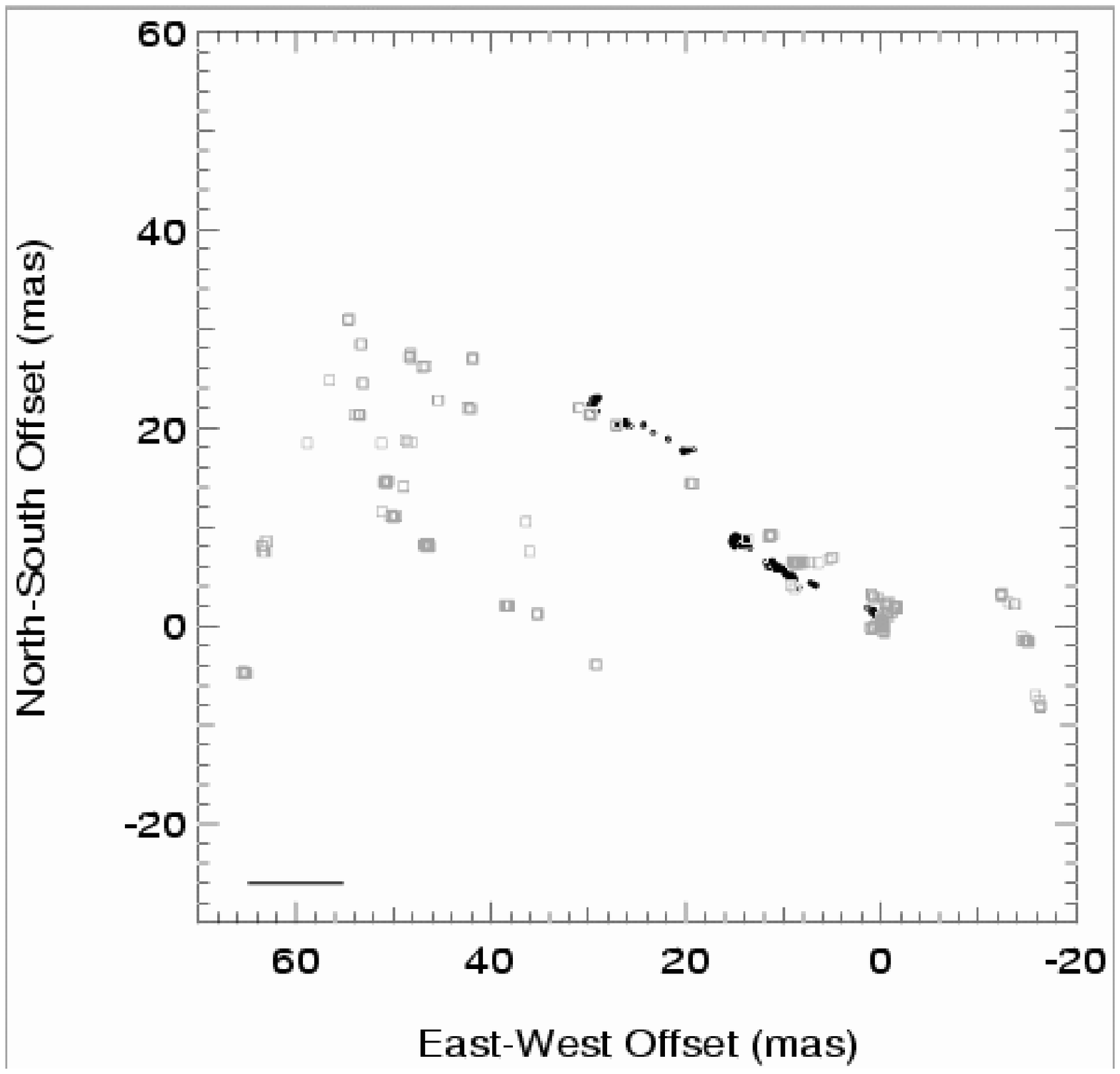]{Map for 1998 June 
where disk masers {\it (black)} are distinguished from outflow masers {\it
(gray)}.  Maser spots that lie more than 1 mas from the midline of the disk are
considered to be outflow masers.  The scale bar corresponds to 0.2\,pc.
\label{disk.vs.wind.map}}

\figcaption[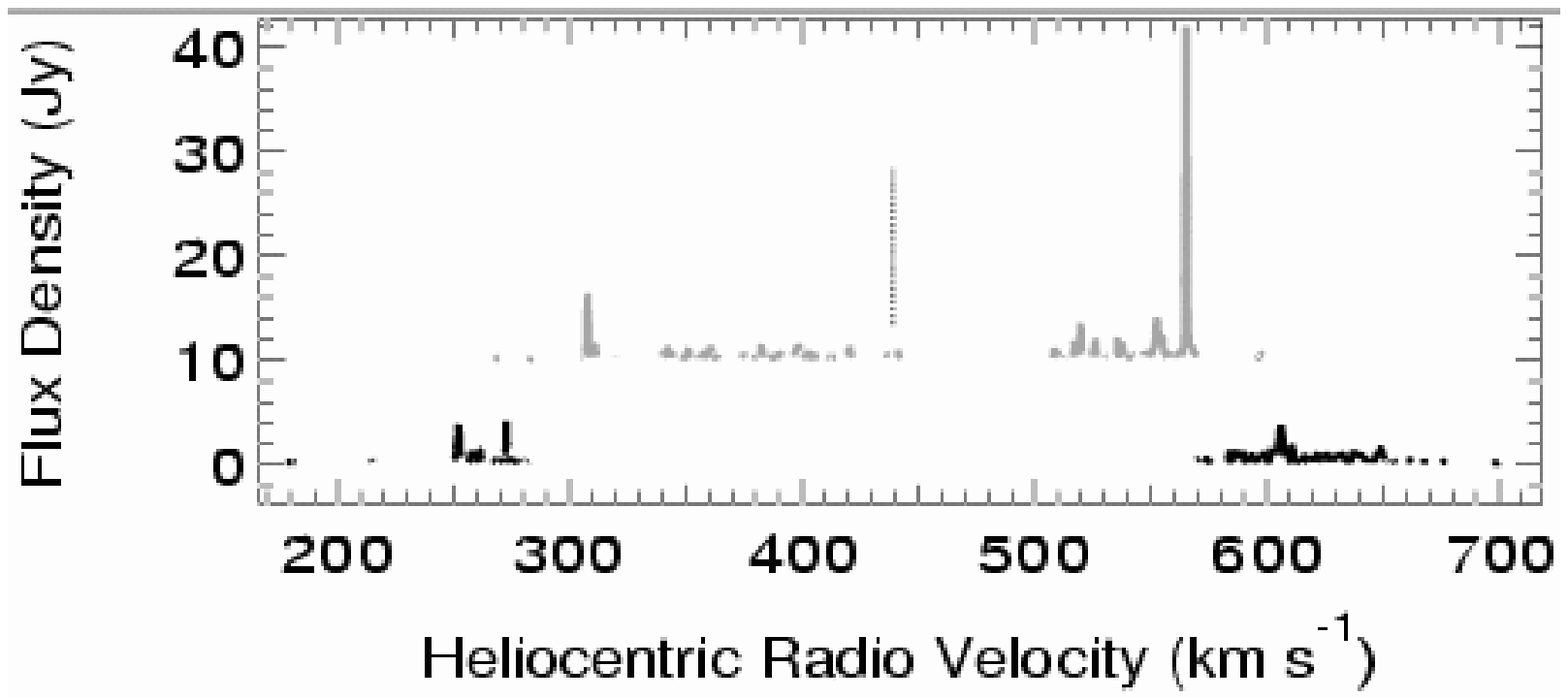]{Total imaged power for 1998 June in which emission from
the disk is shown in black. Other emission is shown in gray and is offset in
flux density for clarity. The vertical line marks the systemic velocity.   The
spectrum for 1998 is shown because it is the best sampled overall. Outflow
emission is offset up to $\sim\pm160$\kms~from the systemic velocity.  
In 1997 July outflow emission was offset up to about $-160$ to $+190$\kms.
\label{disk.vs.wind.spectra}}

\figcaption[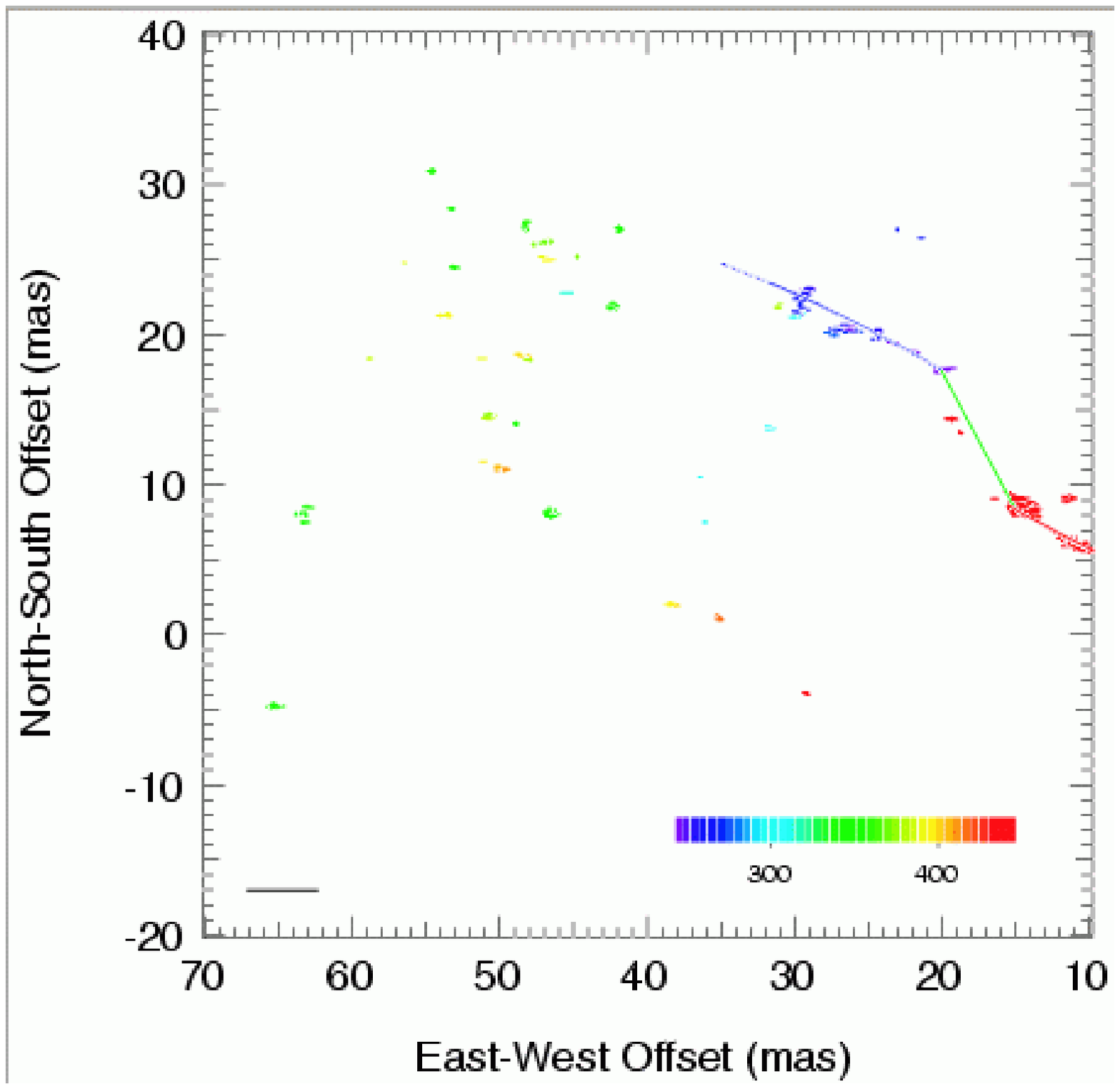]{Enlargement of the sky distribution of the
H$_2$O maser spots associated with the putative outflow.  The color scale is
expanded, as shown by the scale bar, to indicate better the relationship of
line-of-sight velocity and angular structure. The absence of obvious gradients
indicates the emission does not originate in an organized rotating disk, however
extreme the warp may be.  The black scale bar corresponds to 0.2\,pc.
\label{mapzoom}}

\figcaption[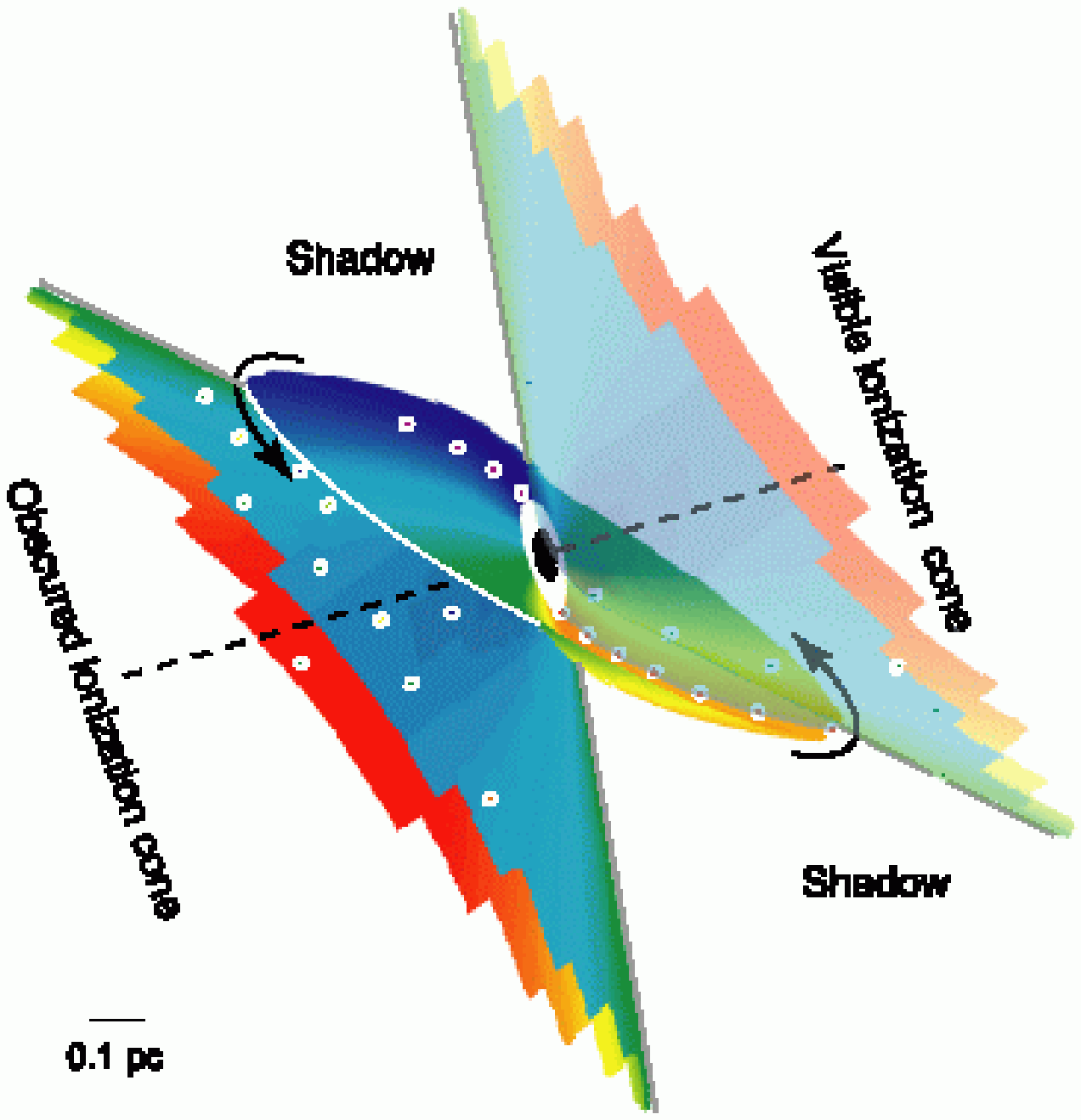]{Model of a thin, warped Keplerian accretion disk and
the surface of a wide-angle bipolar outflow from the central engine.  The disk has
been tipped down at the front for clarity of perspective.  Dots represent maser
spots.  Throughout, color indicates Dopper shift with respect to the systemic
velicity, as in other figures.  The warp of the disk shadows the surroundings, and
maser emission appears only where there is a direct line of sight to the central
engine. The location of the unshadowed region to the west is consistent with the
know optical ionization cone \citep[see, e.g.,][]{vb97}.   The case of Circinus
appears to provide direct evidence that warped disks may determine the extent of
ionization cones in AGN, as well as channel nuclear outflows, at least in some
sources. 
\label{model}}

\newpage

\begin{figure}
\plotone{f1.eps}
\figurenum{1}
\caption{Greenhill\etal}
\end{figure}

\begin{figure}
\plotone{f2.eps}
\figurenum{2}
\caption{Greenhill\etal}
\end{figure}

\begin{figure}
\plotone{f3.eps}
\figurenum{3}
\caption{Greenhill\etal}
\end{figure}

\begin{figure}
\epsscale{0.6}
\plotone{f4.eps}
\figurenum{4}
\caption{Greenhill\etal}
\end{figure}

\begin{figure}[th]
\plotone{f5.eps}
\figurenum{5}
\caption{Greenhill\etal}
\end{figure}

\break

\begin{figure}[th]
\plotone{f6.eps}
\figurenum{6}
\caption{Greenhill\etal}
\end{figure}

\begin{figure}
\plotone{f7.eps}
\figurenum{7}
\caption{Greenhill\etal}
\end{figure}

\begin{figure}
\plotone{f8.eps}
\figurenum{8}
\caption{Greenhill\etal}
\end{figure}

\end{document}